\documentstyle[aps,twoside,epsf]{revtex}

\begin{document}
\draft

\title{Static Color-Coulomb Force}

\author{Attilio Cucchieri\thanks{Electronic address:
         attilioc@asterix.physics.nyu.edu}
and Daniel Zwanziger\thanks{Electronic address:
              zwanzige@acf2.nyu.edu}\thanks{Research supported
in part by the National Science Foundation under grant no.
PHY93-18781}}
\address{{\small\it Physics Department}       \\
  {\small\it New York University}         \\
  {\small\it New York, NY 10003}}

\date{\today}
\maketitle
\begin{abstract}
The static color-Coulomb interaction potential is calculated as the
solution of a non-linear integral equation which
arises in the Hamiltonian Coulomb 
gauge when the restriction to the interior of
the Gribov horizon is implemented.  The potential obtained is in
qualitative agreement with expectations, being Coulombic with logarithmic
corrections at short range, and confining at long range.  The values
obtained for the string tension and $\Lambda_{\overline{MS}}$
are in semi-quantitative
agreement with lattice Monte Carlo and phenomenological determinations.
\end{abstract}

%%%%%%%%%%%%%%%%%%%%%%%%%%%%%%%%%%%%%%%%%%%%%%%%%%%%%%%%%%%%

        The interaction energy $E(r)$ of a pair of heavy quarks at separation
$r$ is a prominent feature of QCD phenomenology.  From fits to the charmonium
and bottonium spectra \cite{Eich,Rich}, $E(r)$ is well-known in the range 
$1\, (\mbox{GeV})^{- 1} < r < 4\, (\mbox{GeV})^{- 1}$.
To calculate $E(r)$ from first principles, one makes a Born-Oppenheimer
approximation, in which heavy quarks move slowly, while gluons and light
quarks rapidly adjust to the instantaneous position of the heavy quarks.
In this approximation, $E(r)$ is the potential energy that appears in a
non-relativistic Schroedinger equation that describes the slow motion of
the heavy quarks, and from which the phenomenological fit for $E(r)$ is
obtained. It is calculated as the gauge-invariant ground-state energy
${\cal H} \Psi_0 = E(r)\Psi_0$, where ${\cal H}$ is the
field-theoretic QCD Hamiltonian that describes the dynamics of gluons and
light quarks in the presence of an external quark-anti-quark pair at fixed
separation $r$.  In the absence of light or dynamical quarks, it is generally
believed that $E(r)$ grows linearly at large $r$, $E(r) \sim Kr$, with
string-tension $K$, and this belief is supported by lattice-gauge
calculations \cite{Sch}.  Pair production of dynamical quarks from the vacuum
causes ``breaking'' of the string at large $r$. However, in the above energy
range where $E(r)$ is known phenomenologically, string-breaking is not yet
manifest, and we shall ignore dynamical quarks in the following.

The Coulomb-gauge is a so-called ``physical'' gauge in which the
conjugate dynamical variables are the {\em 3-dimensionally}
transverse fields $A = A^{tr}$ and $(-E^{tr})$, where
$E_{x,i}^a \,=\, (E^{tr})_{x,i}^a \,-\, (\nabla_i\varphi^a)_x $
and $\varphi$ is the color-Coulomb operator.  Although the Coulomb-gauge is not
known to be renormalizable, a lattice Coulomb-gauge Hamiltonian has been
derived recently from the transfer matrix of Wilson's gauge-invariant
Euclidean lattice gauge theory \cite{Z}, so that gauge-invariant
quantities such as $E(r)$ have the same continuum limit as in Wilson's
theory. For brevity, equations
will be written formally in
continuum notation. 
{\em The exact lattice equations may be found in} \cite{Z}.

The transversality condition on $A$ is not a complete gauge fixing because of
the existence of Gribov copies \cite{Gr}. To obtain a complete gauge
fixing, we adopt the {\em minimal} Coulomb gauge.  In this gauge, the set
$\Lambda$ of gauge-fixed configurations,
called the fundamental modular region (FMR), consists of configurations $A$
for which the Hilbert norm $\|A\|^2 \equiv \int d^3x |A|^2$ of $A$ is minimized
with respect to local gauge transformations $g$:
$\Lambda \equiv \{A:\|A\| \leq \|A^g\| \, {\rm for \, all \,} g \}$. Here
$A^g = g^{\dag} A g + g^{\dag} \nabla g$ is the gauge transform of $A$.  
Configurations with this property are
transverse, $A = A^{tr}$ --- so this gauge falls into the class of Coulomb-gauges ---
and moreover the {\em Faddeev-Popov} (FP)
operator $M(A) \equiv - D\cdot \nabla$
is symmetric $M(A) = M^{\dag} (A)$ and positive $M(A) \geq 0$ for all $A \in \Lambda$
\cite{Z}. Here $D = D(A)$ is the gauge-covariant derivative
$D^{a,c}(A) \equiv \delta^{a,c} \nabla + g_0f^{abc}A^b$. In the following we write
$A$ for $A^{tr}$.

With $E = E^{tr} - \nabla \phi$, the non-Abelian Gauss's law constraint, $D\cdot E =
g_0\rho_{qu}$, may be written $M\phi = g_0\rho$. Here $\rho_{qu}$ is the color-charge
density of the quarks, and $\rho^a \equiv - f^{abc}A^{tr,b}E^{tr,c} + \rho_{qu}^a$
is the color-charge density of the dynamical degrees of freedom. Because $M(A)$ is
a positive operator in the minimal Coulomb gauge, the color-Coulomb field $\phi$ may
be expressed unambiguously in terms of the dynamical degrees of freedom,
$\phi = g_0 M^{-1}\rho$, and so also the color-electric field
$E = E^{tr} - \nabla \phi$. 
For wave functionals $\Phi(A)$ and $\Psi(A)$ defined for
$A$ in $\Lambda$, and with $E^{tr} = i\delta/\delta A^{tr}$, the
Coulomb-gauge Hamiltonian \cite{Lee} is defined by the quadratic form 
\begin{equation}
(\Phi, H_{\rm coul} \Psi) = \int_{\Lambda} dA^{tr} \sigma 
\int d^3x \, 2^{-1}[(E_i^a\Phi)^{*} E_i^a\Psi + \Phi^{*} B_i^a B_i^a\Psi]
\;\mbox{,}
\end{equation}
where $E$ has just been defined, 
$B_i^a \equiv \nabla_jA_k - \nabla_kA_j + f^{abc}A_j^bA_k^c$ for $i, j,$ and $k$ cyclic, 
$\sigma \equiv {\rm det}[M(A)/M(0)]$, and the quark Hamiltonian is suppressed. This
Hamiltonian is symmetric with respect to the inner product 
$(\Phi, \Psi) = \int_{\Lambda} dA^{tr} \sigma\, \Phi^{*} \Psi$, and $\sigma(A)$
is positive because $M(A)$ is.

The main novel ingredient in the present approach is the implementation of the
restriction of the preceding integrals to $\Lambda$, a non-perturbative
effect \cite{FLPNnote}. 
For finite
quantization-volume $V$, the exact boundary of
$\Lambda$ is not known.  However, in \cite{Z} it is argued that, for
periodic boundary conditions and in the limit of large $V$, this region is adequately
approximated by $G/V \leq 0$, and moreover that this restriction may be implemented
by use of the effective Hamiltonian 
$H_{\rm coul} \Rightarrow H_{\rm eff} = H_{\rm coul} + \gamma_0 G$.  Here the
{\em horizon function} $G(A)$ is defined for structure group $SU(N)$ by 
\begin{equation}
G(A) \equiv \int d^3x \, d^3y \, D_{x,i}^{a,c} D_{y,i}^{a,d} (M^{-1})^{c,d}(x,y) 
 - 3 ( N^2 - 1 ) V
\;\mbox{.}
\end{equation}
The term $\gamma_0 G$ makes the wave-functionals in the
Fock space of $H_{\rm eff}$ vanish rapidly outside $\Lambda$, so the restriction on
the integrals may be ignored, $\int_{\Lambda} dA^{tr} \Rightarrow \int dA^{tr}$.
Here $\gamma_0$ is a thermodynamic parameter that sets the scale for hadronic masses. 
Its value is determined by the {\em horizon condition} $\langle G\rangle /V$ = 0.
Here and below, the expectation-value is taken in the ground state of $H_{\rm eff}$.
The horizon condition expresses the fact that, for large $V$, the probability gets
concentrated on the boundary $G/V = 0$.

The Coulomb energy
\begin{equation}
E_{\rm coul} = \int dA^{tr} \, \sigma
\int d^3x \, 2^{-1} | g_0 \nabla (M^{-1} \rho) \Phi |^2
\end{equation}
contributes additively to the energy
of a state $\Phi$, the remaining terms being positive.
Consider the contribution
to $E_{\rm coul}$ from the part of
$M^{-1}$ which is diagonal in momentum space, 
\begin{equation}
V^{-1}(M^{-1})_{q,a:q,b} =
\int d^3y \, e^{iq\cdot y}\, [V^{-1} \int d^3x \, (M^{-1})_{x,a;x+y,b}]
\;\mbox{.}
\end{equation}
For each $y$,
the term in brackets is a bulk quantity per unit volume. By translation invariance, its
covariance matrix is of order $V^{-1}$.  Consequently, for large $V$, it approaches its
mean-field value
\begin{equation}
V^{-1} (M^{-1})_{q,a;q,b} \, = \, C(q) \delta^{a,b} + {\cal O}(V^{-1/2})
\;\mbox{,}
\end{equation}
where
\begin{equation}
C(q) \delta^{a,b} \,=\, \int d^3y \, e^{iq\cdot y} \langle (M^{-1})_{x,a;x+y,b}
\rangle
\end{equation}
is the 3-dimensional 
FP propagator. Thus
$\phi$ and $E_{\rm coul}$ receive the contributions 
$\tilde{\phi}(q) = g_0 C(q) \tilde{\rho}(q)$ and 
\begin{equation}
\int dA^{tr} \sigma \, (2V)^{-1}$ $\sum_q q^2 |g_0 C(q) \tilde{\rho}(q) \Phi |^2
\;\mbox{,}
\end{equation}
where the color-charge density operator $\tilde{\rho}(q)$ satisfies 
${\rm lim}_{q\rightarrow 0}\tilde{\rho}^a(q) = Q^a$. Here $Q^a$, with 
$[Q^a, Q^b] = f^{abc}Q^c$, is the total color-charge.

The FP propagator is of the form
$C^{-1}(q) = q^2 - q_i \Sigma_{i,j} q_j$, where $\Sigma_{i,j}(q)$ is a self energy.
The horizon condition is equivalent \cite{Z} to the condition
$\Sigma_{i,j}(0) = \delta_{i,j}$. This gives 
$C^{-1}(q) = q_i[\Sigma_{i,j}(0) - \Sigma_{i,j}(q)]q_j$, so $C(q)$ is more
singular at $q = 0$ than $q^{-2}$. We conclude that the restriction to the FMR
produces a long-range color-Coulomb field, as originally foreseen by Gribov
\cite{Gr}.  

The self-energy $\Sigma$ satisfies a Schwinger-Dyson equation which we write
symbolically 
$ q \,\Sigma \, q = R(q) \equiv g_0^2 \int \Gamma_0 \, D \, C \, \Gamma$, which gives 
$C^{-1}(q) = R(0) - R(q)$. Here $D$ is the gluon propagator, and $\Gamma$
and $\Gamma_0$ are the exact and zeroth-order ghost-ghost-gluon vertex functions.
When $R$ is evaluated in the
ground state of $H_{\rm eff}$, this equation \cite{Zfoot}
provides the mean-field
self-consistency condition (MSC) that determines the mean-field {\em function} $C(q)$.   

So far our results are exact, but in the following we resort to
approximations to solve the MSC. This equation contradicts
the usual perturbative expansion, as one sees from the appearance of
$(g_0)^2$ in $R$, whereas $C, D$, and $\Gamma$ are nominally of
leading order $(g_0)^0$. To obtain an expansion
in powers of $g_0$ that is consistent with the MSC, we assume that $C(q)$ is of leading
order $(g_0)^{-1}$, namely $C(q) = g_0^{-1}u(q) + O(1)$, and that all other correlation
functions are analytic in $g_0$. To leading order, the MSC is an integral equation
that determines $u(q)$:   
\begin{equation}
u^{-1}(q)\, =\, N (2\pi)^{-3} \int d^3k\, D_0(k)[q^2 - (q\cdot k)^2/k^2]
[u(k) - u(k-q)]
\;\mbox{.}
\label{eq:intu}
\end{equation}
The kernel $D_0$ is the gluon propagator
\begin{equation}
D_0(k) (\delta_{i,j} - k_i k_j/k^2) \delta^{b,c} = \int d^3x \, e^{-ik\cdot x} 
\langle A_{x+y,i}^b A_{y,j}^c \rangle_0 
\;\mbox{,}
\end{equation}
evaluated in the ground state of 
\begin{equation}
H_0 \equiv \sum_{k, \lambda} \omega_k \, a_{k, \lambda}^{\dag} a_{k, \lambda}  +
(2V)^{-1} \sum_k \tilde{\rho}_{-k}^a v_k \tilde{\rho}_{k}^a
\;\mbox{,}
\end{equation}
where $a_{k, \lambda}^{\dag}$ and $a_{k, \lambda}$ are creation
and annihilation operators for
$A^{tr}$ and $E^{tr}$, $\lambda$ is a two-valued polarization index, and 
$\omega_k = (k^2 + \mu^4 u_k)^{1/2}$.  This is the zeroth-order
part of $H_{\rm eff}$, obtained by systematically expanding $H_{\rm eff}$ in powers of
$g_0$, with $\gamma_0$ scaled according to $\gamma_0 = (2Ng_0)^{-1}\mu^4$.
In addition to a harmonic oscillator Hamiltonian, 
$H_0$  also contains a color-Coulomb interaction
Hamiltonian with interaction potential $v_k \equiv k^2 u_k^2$ that is
independent of $g_0$, as is consistent with dimensional transmutation.

This term prevents us from calculating $D_0(k)$ exactly. 
We neglect it, so $D_0$ is approximated by
$D_0^{(0)}(k) \equiv (2\omega_k)^{-1}$, and we call
$w_q \equiv u_q^{(0)}$ the solution to (\ref{eq:intu}) with kernel $D_0^{(0)}$. 
We expect that $D_0(k)$ is even
more suppressed at low $k$ than $D_0^{(0)}(k)$, 
so by (\ref{eq:intu}) $u_q$ is enhanced at low $q$ compared to $w_q$
corresponding to an even longer range force than we find.
 
The asymptotic form of $w_q$ at high and low $q$
has been determined analytically \cite{Z}, and it has been found
numerically for intermediate $q$ with $2 \%$ accuracy
(see \cite{Athesis,CZ} for details).
We write $w(q) = g_c(q)/q^2$, where $g_c(q)$ is a
renormalized running Coulomb-coupling constant, and we express our results in
terms of the static color-Coulomb potential
$v_q = q^2 w^2(q) = g_c^2(q)/q^2$, which appears
in $H_0$, and the Fourier transform of $v_q$ (normalized for a pair of external quarks),
which is given (apart from an additive constant) by 
\begin{equation}
V(r) = - (N^2 - 1)
(4 N \pi^2)^{-1} \int_0^{\infty} dq \, g_c^2(q) \,(qr)^{-1} \, \sin(qr)
\;\mbox{,}
\end{equation}
corresponding to a force $f(r) = -V'(r)$.  

The asymptotic form of $g_c$ at high $q$ is given by
\begin{equation}
g_c^{-2} \, = \, b_c \, t + 6^{-1} b_c \, \ln(b_c t) + {\cal O}(t^{-1} \ln t)
\;\mbox{,}
\end{equation}
where $t = {\rm ln}(q/m)$, $b_c = (3\pi^2)^{-1}N$, $m = C \mu$, and $C$ is a constant. 
The first and second terms are of the form of the one- and two-loop contributions to the
running coupling constant $g_{\rm rg}$ of the perturbative renormalization group, but the
coefficients are different. This difference arises because at high momentum the
interaction is not purely static Coulombic, and it is verified in \cite{Z}
that the difference in the first coefficient is correctly accounted for by terms
that are
neglected here. The coefficient of the second term is a new result. The limiting
behavior of the force $f(r)$ at small $r$ is given by 
\begin{equation}
f(r) \approx 3 \pi (N^2 - 1) (4 N^2 r^2)^{-1} [\log(\Lambda_R^2 r^2) + 3^{-1}
\log \log(\lambda^2 r^2)]^{-1}
\;\mbox{,}
\label{eq:fsmallr}
\end{equation}
where $\Lambda_R \equiv e^{\gamma -1}m$,
$\gamma$ is Euler's constant, and $\lambda$ is a constant.

The confinement properties of the theory in the present approximation are determined by
the asymptotic form of $g_c(q)$ at low
$q$, which is given by
$g_c(q) = B(\mu/q)^{4/3}$, where $B^{-3/2} = N \pi^{-2} \Gamma(8/3) \Gamma(2/3) /
\Gamma(16/3)$ (see \cite{Z}).
This corresponds to
a color-Coulomb potential $V(r) \sim r^{5/3}$ which rises
more rapidly with $r$ than a string tension $Kr$.
This somewhat surprising result may be an
artifact of the approximations made. On the other hand $V(r)$, which appears in the
(approximate) quantum-field theoretic Coulomb-gauge Hamiltonian, must be distinguished
from the gauge-independent quark-pair energy $E(r)$ discussed in the introduction,
and it is shown in \cite{Z} that $E(r) \leq V(r)$.
If $V(r)$ does grow more rapidly than $Kr$ at large $r$, then the wave-function
$\Psi(A^{tr})$, whose
defining property is to minimize $E(r)$, adjusts itself so $E(r)$ rises no more
rapidly than $Kr$ \cite{Seiler}.
It does so by changing the two superposed spherically symmetric long range color-Coulomb
fields into a flux tube.

We shall 
compare our results for $V(r)$ with phenomenological fits
to $E(r)$, to see if there is a range of ``small'' $r$ for which $V(r)$
agrees with
$E(r)$, as suggested by asymptotic freedom. To this end,
we consider
two phenomenological models:
the Cornell potential \cite{Eich} and the
Richardson potential \cite{Rich}, both of 
which fit the $c {\overline c}$ and $b {\overline b}$
spectra well.

To connect dimensionless
quantities and the real world, we fix the length scale
by using Sommer's \cite{Sommer}
dimensionless phenomenological relation
$r^{2}_{0}\,f(r_{0}) = - 1.65$
which holds for the Cornell force at
$r_{0}\,a \equiv R_{0}\,=\,2.48 \,(\mbox{GeV})^{- 1}$.
From the value of $a$ in $(\mbox{GeV})^{- 1}$ we obtain
the string tension
$\sqrt{\,\sigma\,} \equiv \sqrt{\min_{r}\left[\,- f(r) \,\right]}\,a^{- 1}
\,=\,518\,\mbox{MeV}$.
[By this definition, we are
evaluating $\sigma$ where $f^{'}(r) = 0$, {\em i.e}\ where the
potential is approximately linear.]
It is not easy to estimate an uncertainty for the string tension.
However, its value seems to depend very weakly on the values
of the parameters of our trial solution \cite{Athesis}.
If we identify the parameter $\Lambda_{R} \equiv e^{\gamma -1}m$ in
(\ref{eq:fsmallr}) with the corresponding physical parameter \cite{Billoire},
then from the relation \cite{Billoire}
$\Lambda_{\overline{MS}} \,=\, \Lambda_{R} \, \exp{( - \gamma + 1 - 31 / 66 )} \,$
we obtain $\Lambda_{\overline{MS}} = 124\,\pm\,12\,\mbox{MeV}$
(see \cite{Athesis,Afoot}).
In Figure \ref{FIG:plot_4forze}a we plot our result for
$f(r)$, the two Cornell forces \cite{Eich}, and the Richardson force \cite{Rich}.
Our force gets its maximum value at a separation
of about $2.5 (\mbox{GeV})^{- 1}$ and it is almost constant
up to $4 (\mbox{GeV})^{- 1}$, the variation being of order
$12 \%$.

These results depend on the
phenomenological conditions that
we have used to set the length scale.
If we use the dimensionless relation $r^{2}_{0}\,f(r_{0}) \,=\, - 1.35$,
which holds for the Richardson force at
$R_{0} = 2.48 (\mbox{GeV})^{- 1}$,
we obtain
$\sqrt{\,\sigma\,} = 468\,\mbox{MeV}$,
$\Lambda_{\overline{MS}} = 118\,\pm\,12\,\mbox{MeV}$
and the plot shown in Figure \ref{FIG:plot_4forze}b.
In this case the agreement
is even better:
our force reaches its maximum value at a separation
of about $2.75 (\mbox{GeV})^{- 1}$, and its variation
at $4 (\mbox{GeV})^{- 1}$ is of order $8 \%$.

As an exact result, we have found that the restriction to the
fundamental modular region causes an infrared singularity of the
color-Coulomb propagator and thus a long-range color-Coulomb potential.
Although our calculation of this potential required possibly
severe approximations that are describe above, nevertheless it gives
results which are in qualitative agreement with phenomenologically
determined potentials.
The fit is surprisingly good.
After setting the scale by Sommer's method,
the force is in qualitative agreement
with phenomenological models, and
the values obtained for the string
tension and $\Lambda_{\overline{MS}}$ are in semi-quantitative agreement
with lattice Monte Carlo and phenomenological
determinations (see \cite{Eich,Sch} and \cite{Particle}).
It would appear that the approximate equality
$V(r) \approx E(r)$ extends to the range $r < 4 (\mbox{GeV})^{- 1}$,
and moreover that the approximations made in our calculation of
$V(r)$ do not qualitatively destroy this agreement. Although there is no
{\em a priori}
reason to expect that vacuum polarization of gluons should not be
important in this range, this may not be so surprising after all.
For although
vacuum polarization (pair production) of quarks does
``break'' the string,
this is not yet manifest for $r < 4 (\mbox{GeV})^{- 1}$.

We are grateful to Tony Duncan, Martin Schaden and Alberto
Sirlin for informative discussions.

%%%%%%%%%%%%%%%%%%%%%%%%%%%%%%%%%%%%%%%%%%%%%%%%%%%%%%%%%%%%%%%%%%%%%%%%

\begin{figure}[t]
\begin{center}
\vspace*{0cm} \hspace*{-0cm}
\epsfxsize=0.4\textwidth
\leavevmode\epsffile{figure1a.ps}
\hspace{2cm}
\epsfxsize=0.4\textwidth
\epsffile{figure1b.ps}  \\
\caption{~Plot of: (i) our force $f(R)$
         (the curve which is increasing negative
         at large $R$), (ii)
         the forces (for the $c {\overline c}$ and the
         $b {\overline b}$ cases)
         derived from the Cornell potential (the two curves
         very close to each other), and (iii)
         the Richardson
         force (the highest curve).
         In case $1$ we set the length scale
         by $r^{2}_{0} \, f(r_{0}) = - 1.65$
         from the Cornell force,
         while in case $2$ we use the
         condition $r^{2}_{0} \, f(r_{0}) = - 1.35$ from the
         Richardson force,
         both at $R_{0} = 2.48 (\mbox{GeV})^{- 1}$.}
\label{FIG:plot_4forze}
\end{center}
\end{figure}


\begin{references}
 
\bibitem{Eich} E.\ Eichten et al.,
               Phys.\ Rev.\ Lett. {\bf 34} (1975) 369;
               E.\ Eichten et al.,
               Phys.\ Rev.\  {\bf D21} (1980) 203;
               E.\ Eichten and F.\ Feinberg, Phys.\ Rev.\  {\bf D23}
               (1981) 2724.

\bibitem{Rich} J.\ L.\ Richardson, Phys.\ Lett.\  {\bf B82} (1979) 272.

\bibitem{Sch} G.\ S.\ Bali and K.\ Schilling, Nucl.\ Phys.\  {\bf B} (Proc. Suppl.)
              {\bf 34} (1994) 147.

\bibitem{Z} D.\ Zwanziger, Nucl.Phys.\ {\bf B485} (1997) 185.

\bibitem{Gr} V.\ N.\ Gribov, Nucl.\ Phys.\ {\bf B139} (1978) 1.

\bibitem{Lee} N.\ Christ and T.\ D.\ Lee, Phys.\ Rev.\ {\bf D22} (1980) 939.

\bibitem{FLPNnote} An alternative approach which has been advocated recently is
               to integrate over all $A^{tr}$ without restriction, but with
               $\sigma(A)$ as a signed measure [see 
               R.\ Friedberg et al., 
               Ann. of Phys. {\bf 246} (1996) 381].

\bibitem{Zfoot} In the Landau gauge, where
                $\tilde{Z}_1' = 1$, this equation is invariant under
                renormalization --- the Z's cancel! ---
                and a corresponding property should also hold in the Coulomb-gauge.

\bibitem{Athesis} A.\ Cucchieri, {\em Numerical Results in Minimal Lattice
      Coulomb and Landau Gauges: Color-Coulomb Potential and Gluon and Ghost
      Propagators}, PhD thesis, New York University (May 1996).

\bibitem{CZ} A.\ Cucchieri and D.\ Zwanziger, 
             Nucl.Phys.\ {\bf B} (Proc.\ Suppl.) {\bf 53} (1997) 815.
             
\bibitem{Seiler} E.\ Seiler, Phys.\ Rev.\ {\bf D22} (1980) 2412.   

\bibitem{Sommer} R.\ Sommer, Nucl.\ Phys.\  {\bf B411} (1994) 839.

\bibitem{Billoire} A.\ Billoire, Phys.\ Lett.\  {\bf B104} (1981) 472.

\bibitem{Afoot} The uncertainty on $\Lambda_{\overline{MS}}$ comes from
                the uncertainties on the values of $a$, and 
                of the parameter $m$ entering
                into the definition of $\Lambda_{R}$.
                By solving equation (\ref{eq:intu})
                numerically, we have found the value of $m$ with an
                estimated accuracy of about $10 \%$. We have 
                considered negligible the uncertainty
                on $a$.

\bibitem{Particle} Particle Data Group,
                   Review of Particle Properties,
                   Phys.\ Rev.\  {\bf D50} (1994) 1173.

\end{references}
\end{document}